\documentclass{article}
\usepackage{spconf,amsmath,graphicx}
\usepackage {booktabs}
\usepackage{hyperref}
\hypersetup{
    colorlinks=true,
    citecolor=blue,
    linkcolor=black,
    linkbordercolor={1 1 1},
    citebordercolor={1 1 1},
}
\usepackage[export]{adjustbox}
\usepackage{caption,subcaption, floatrow}

\title{An Empirical Study of Visual Features for DNN based Audio-Visual Speech Enhancement in Multi-talker Environments}
\name{Shrishti Saha Shetu, Soumitro Chakrabarty and Emanu\"{e}l A. P. Habets}
\address{Fraunhofer IIS, Am Wolfsmantel 33, 91058 Erlangen, Germany \\
        shetu.nitjsr13@gmail.com,    
        \{soumitro.chakrabarty, emanuel.habets\} @iis.fraunhofer.de}
\begin{document}
\maketitle
\begin{abstract}
Audio-visual speech enhancement (AVSE) methods use both audio and visual features for the task of speech enhancement and the use of visual features has been shown to be particularly effective in multi-speaker scenarios. In the majority of deep neural network (DNN) based AVSE methods, the audio and visual data are first processed separately using different sub-networks, and then the learned features are fused to utilize the information from both modalities. There have been various studies on suitable audio input features and network architectures, however, to the best of our knowledge, there is no published study that has investigated which visual features are best suited for this specific task. In this work, we perform an empirical study of the most commonly used visual features for DNN based AVSE, the pre-processing requirements for each of these features, and investigate their influence on the performance. Our study shows that despite the overall better performance of embedding-based features, their computationally intensive pre-processing make their use difficult in low resource systems. For such systems, optical flow or raw pixels-based features might be better suited.   
\end{abstract}
\section{Introduction}
\label{sec:intro}
The aims of the speech enhancement (SE) algorithms are to improve the quality and, if possible, also the intelligibility of a noisy speech signal. Traditional SE algorithms only use audio features and assume the background interference is non-speech. On contrary, in multi-talker environments (cocktail party effect), where more than one speakers are talking simultaneously, the audio only (AO) approach cannot enhance the speech of the target speaker without any prior knowledge or special microphone configurations \cite{6853713}. In these circumstances, a joint audio-visual (AV) method can be helpful and enhance the visually present speaker based on the visual features.

Several studies have shown the benefits of using a joint AV model for different tasks. In \cite{golumbic2013visual}, researchers have stated that viewing a speaker’s face significantly enhances a person’s capacity to understand the speech in a noisy environment. The use of visual modality has also been proved fruitful in different speech processing algorithms, such as audio visual speech recognition \cite{noda2015audio}, lip reading \cite{Chung16, assael2016lipnet}, and lip to speech synthesis \cite{prajwal2020learning}, etc. Recent studies also demonstrated that the use of visual features can help in speech denoising in very low signal to noise ratio (SNR) conditions \cite{chuang2020lite,hou2018audio}.

In recent years, many research groups have contributed significantly to AVSE. The AVSE methods generally use different types of visual features for processing visual information before fusing with audio features for isolating the target speaker. In \cite{ephrat2018looking}, researchers used face embedding as the visual features, which are actually achieved by a pre-trained face recognition model. Other successful methods have used lip embedding \cite{afouras2018conversation} from a visual speech recognition (VSR) system. Other common visual features are raw pixels, optical flow, or facial landmark-based features \cite{michelsanti2020overview}. 

To foster the research in this field, we believe a collective study of different visual features is highly beneficial, which, to the best of our knowledge, has not been performed yet. Therefore, we present an empirical study of four different groups of visual features. We investigate all these visual features using a joint AV multi-modal model, which incorporates a late fusion strategy. Our investigation consists of experiments with both ideal (less movements, homogeneous background, clean audio and high video quality) and realistic  video recordings. Our study mainly concentrates on the performance of AVSE systems using different visual features. We also investigate the challenge of using different visual features, the computational complexities involved in pre-processing steps and the feasibility to use those visual features in low resource systems.




\section{Neural Network Architecture}
\label{sec:NNA}

 We use the AV multi-modal model architecture proposed in \cite{ephrat2018looking} as the baseline model which was originally proposed for embedding-based visual features. This model has two separate audio and visual streams for processing the input audio and visual features. The learned features from these two streams are fused at a later stage, and processed further.  

The visual stream of this model takes face embedding as input features, and consists of six 2D dilated convolution layers. The spatial convolutions and dilations in the visual stream are performed over the temporal axis. 

The audio stream of this model takes the power law compressed short-time Fourier transform (STFT) as input with the real and imaginary as two channels with compression factor 0.3 (\(\text{A}^{0.3}\), where A is the real or imaginary part of the input/output spectogram). This audio stream consists of fifteen 2D dilated convolutions layers. 
The feature maps of both AV streams are concatenated, and then processed by a bidirectional(Bi)-LSTM followed by three fully connected layers. The output of this model is a complex time-frequency mask. The enhanced spectogram is reconstructed by the complex multiplication between the complex mask and the noisy speech. The mean square error loss between the true and the enhanced spectogram is used as the objective loss function to train the model. 

In all our experiments, the audio stream of this model, the AV fusion strategy and the input and target audio features are always kept constant.

\section{Visual Features}
\label{sec:VFE}


Our study consists of four different groups of visual features, namely, raw pixels, embedding, optical flow and facial landmark-based features. For each of these groups, we extract features for both the face and mouth/lip region.

\subsection{Raw Pixels-based Features}
\label{ssec:Raw}
The normalized raw pixels of extracted face or lip region (can be defined as raw face or lip pixels) are generally used as raw pixels-based features. For lip region extraction, we use facial landmarks detection method presented in \cite{king2009dlib}. While this approach does sufficiently well for detecting lip region, the extracted face region using this method are not consistent and have jumps between consecutive frames. As the raw pixels of the extracted face regions are used as the primary features for face pixels-based AV model (AV-FacePixels) and also further processed for other higher level visual features, the reliability and consistency of these features are highly desirable. Therefore, we extract the face region using the MTCNN model \cite{zhang2016joint}. 

\subsection{Embedding-based Features}
\label{ssec:Emb}
In literature, the embedding-based visual features commonly refer to the facial recognition and VSR embeddings \cite{michelsanti2020overview}.  While these visual features come from two completely different research fields, in recent time, both proved to be effective visual features for AVSE \cite{ephrat2018looking,afouras2018conversation}.

The facial recognition embedding (face embedding)-based features are generally achieved using a face recognition model. The output from one of the last flattened layer (the layer contains no channels) is used as the face embedding.  These face embeddings represent a higher dimensional projection of the input features (raw face pixels) used for recognizing faces, by discarding the irrelevant information such
as illumination. In our experiment, the face embeddings are extracted using an implementation\footnote{\url{https://github.com/nyoki-mtl/keras-facenet}} of Facenet model \cite{schroff2015facenet}. 

Similar to the face embedding, the VSR embedding (popularly known as lip embedding, as most VSR systems use raw lip pixels as visual features) are obtained using a VSR system. We extract two different kind of lip embeddings for our experiments.  The sentence level lip embeddings are extracted using a Lipnet \cite{assael2016lipnet} pre-trained model\footnote{\url{https://github.com/rizkiarm/LipNet}}. The word level lip embeddings are obtained using a temporal convolution network (TCN)-based \cite{martinez2020lipreading} pre-trained model\footnote{\url{https://github.com/mpc001/Lipreading_using_Temporal_Convolutional_Networks}}.

\subsection{Optical Flow-based Features}
\label{ssec:Opt}
Optical flow is defined as the motion of objects between consecutive frames of a sequence caused by the relative movement between the object and camera. There are many variants of optical flow estimation but the most prominent one is the dense optical flow, which estimate the motion or, displacement field from only two frames and try to compensate for the background motion \cite{farneback2003two}. The dense optical flow has been widely used in many research fields along with AVSE systems \cite{simonyan2014two}. In our implementation, for extracting optical flow-based features, we use OpenCV dense optical flow algorithm \cite{baker2004opencv}. The two spatial directional optical flows are calculated, from which the direction and magnitude are extracted. Finally, a three dimensional feature is obtained using color coding.

\subsection{Facial Landmark-based Features}
\label{ssec:landmark}
Landmark-based motion features use the relative motion between two consecutive frames, as described in \cite{morrone2019face}. We extract these features by simply taking the first derivatives between the extracted grayscale face and lip region of consecutive frames. Compared to optical flow-based features, landmark-based motion features do not compensate for the background motion and, therefore, can be vulnerable to rapid movements between consecutive frames.

\subsection{Integration of Visual Features}
\label{ssec:Inte}
To incorporate all the above mentioned visual features (except embedding-based features), we also modify the visual stream of the baseline embedding-based AV model, as this model architecture can not process 3D raw pixels-based or optical flow-based features. We use a Lipnet architecture for the visual stream of this non-embedding-based AV model, as proposed in \cite{assael2016lipnet}. The Lipnet model consists of 3\(\times\)(spatiotemporal convolutions, channel-wise dropout, spatial max-pooling), followed by two Bi-gated recurrent units (Bi-GRUs) for processing the visual features. The performance of this architecture in a VSR system using raw lip pixels-based features motivated us to use it in our architecture. For face-based non-embedding features, we also use the same architecture, as a similar network was proven to be successful in \cite{gabbay2018seeing}. However, to keep the number of parameters almost similar to those of the embedding-based architectures, we use only one Bi-GRU. 


 


\begin{table}[t]
\footnotesize
\centering
\begin{tabular}{lcc}
\toprule
\textbf{Model Name}  & \textbf{\begin{tabular}{@{}c@{}}Visual\\ Features\end{tabular}} & \textbf{Input Shape}\\
\midrule 
AV-Faceembs  & Face Embedding& (1,1792)\\
AV-Lipembs &  Lip Embedding& (1,512)\\
AV-Lipembs-LRW & Lip Embedding& (1,512)\\
AV-FacePixels  & Raw Face Pixels& (100,100,3)\\
AV-LipPixels&Raw Lip Pixels&(50,100,3)\\
AV-FaceOpt&  Opt. Flow of Faces& (100,100,3) \\
AV-LipOpt&   Opt. Flow of Lips &(50,100,3)\\
AV-FaceMotion&  Face Motion &(100,100,1)\\
AV-LipMotion&  Lip Motion&(50,100,1)\\
\bottomrule
\end{tabular}
\caption{Specifications of visual features used in different experimental models.}
\label{fig:Visuals}
\vspace{-2mm}
\end{table}
\vspace{-4mm}




\section{Experiments}
\label{sec:Experiments}

\subsection{Datasets}
\textbf{\textit{TCD-TIMIT Dataset: }}For our experiments in ideal condition, we use TCD-TIMIT dataset \cite{harte2015tcd}. This dataset consists of 59 speakers with around 160-200 videos each and also sentences from three lipspeakers. The videos have a homogeneous background and the consecutive frames contain very little movements, which are suitable for extracting highly reliable visual features.\\
\\\textbf{\textit{LRS3 Dataset: }}For our experiment with real-life condition, we use the LRS3 dataset \cite{afouras2018lrs3}, which consists TED and TEDx videos of almost 5000 speakers. The video resolution of this dataset is significantly lower than TCD-TIMIT dataset, with varying background, and speaker movement.

\subsection{Experimental Setup}
\label{ssec:expsetup}

In total, we trained 9 different AV models using the above mentioned visual features. In Table \ref{fig:Visuals}, the input shape of different visual features and corresponding model names are shown. In case of  the two lip embedding-based models, a selection is made based on the training dataset. For TCD-TIMIT case, we only train the AV-Lipembs model, as the pretrained network for extracting embeddings for this model was trained with GRID Corpus dataset \cite{cooke2006audio}, which also corresponds to the same phoneme set as the TCD-TIMIT dataset. In case of LRS3 dataset, we train both AV-Lipembs and AV-Lipembs-LRW models. The LRW dataset \cite{Chung16} (which has significantly larger vocabulary set than GRID corpus dataset) was used to train the underlying TCN-based pre-trained model used for computing the lip embeddings for AV-Lipembs-LRW model.

For comparing the improvement using visual features with the corresponding AO approach, we also developed an AO model. This model is directly adapted from the audio stream of the embedding-based model, as described in Section \ref{sec:NNA}. In our study, all the considered models take 3~s audio segment as input with 16 kHz sampling frequency, which corresponds to 75 visual frames using 25 FPS sampling frequency. For pre-processing the audio features, we use discrete Fourier transform length of 512 with window length of 25 ms and 10 ms hop length.

For the experiment with TCD-TIMIT dataset, we use 52 speakers and first 85\(\%\) sentences of every speaker for training. For creating the noisy mixture, we use the random sentences from GRID Corpus \cite{cooke2006audio}, DNS Challenge Corpora \cite{reddy2020interspeech} and TCD-TIMIT itself. All together we create a training dataset of almost 130 hours.

For evaluation, we use 8 unseen speakers, and rest 15\(\%\) unseen sentences for every speakers. The noisy mixtures were created mixing randomly selected sentences from TCD-TIMIT dataset with the reference sentences. The size of the test dataset was almost 1.2 hours.

The experimental setup with LRS3 dataset consists of 1500 unique speakers. All together 30000 clean sentences (each of 3~s) were used for training. The noisy mixture was created using sentences from LRS3 dataset, and DNS challenge Corpora. The training dataset is of 250 hours. For evaluation, we use almost 250 unseen speakers, over 3000 unseen sentences and interference from both LRS3 and DNS challenge Corpora. The evaluation dataset is of almost 5.5 hours.

In all the experiments, the training mixtures have a uniform SNR distribution between -5 dB to 20 dB. The test mixtures represent a uniform SNR distribution between -5 to 5 dB.

\begin{table*}
\footnotesize
          \centering
          \begin{subtable}{0.4\linewidth}
          \footnotesize
              \centering
        \begin{tabular}{lcc}
        \toprule
        \textbf{Model Name} &  \textbf{$\Delta$SI-SDR (dB)} & \textbf{$\Delta$SDR (dB)} \\
         \midrule 
        AO-Model &  -9.84$\pm$14.12 & 1.30$\pm$3.71  \\
        AV-Faceembs &  8.92$\pm$6.42 & 8.36$\pm$4.26 \\
        AV-Lipembs &  \textbf{9.58$\pm$5.65} & \textbf{8.74$\pm$3.87} \\
        AV-FacePixels &  7.92$\pm$6.52 & 7.79$\pm$4.44 \\
        AV-LipPixels&  6.98$\pm$7.52 & 7.30$\pm$4.74 \\
        AV-FaceOpt  & 9.06$\pm$5.66 & 8.41$\pm$3.96\\
        AV-LipOpt&   7.88$\pm$6.81& 7.83$\pm$4.36\\
        AV-FaceMotion&  6.52$\pm$7.84 & 7.11$\pm$4.64  \\
        AV-LipMotion& 7.78$\pm$6.96& 7.74$\pm$4.54\\
        \bottomrule

        \end{tabular}
        \vspace{-1mm}
              \caption{}
              \label{fig:Models2}
          \end{subtable}%
          \begin{subtable}{.6\linewidth}
          \footnotesize
              \centering
                \begin{tabular}{lcccc}
                \toprule 
                \textbf{Model Name}& \textbf{$\Delta$SI-SDR (dB)}  & \textbf{$\Delta$SDR (dB)}& \textbf{$\Delta$PESQ} & \textbf{$\Delta$STOI} \\
                \midrule 
                
                AV-Faceembs & 8.84$\pm$3.22&  8.19$\pm$2.68& \textbf{0.34$\pm$0.26} & 0.129$\pm$0.082\\
                AV-Lipembs &  7.81$\pm$4.19 &7.49$\pm$2.92 &0.27$\pm$0.25& 0.107$\pm$0.105 \\
                AV-Lipembs-LRW &  8.82$\pm$3.37&8.19$\pm$2.76& \textbf{0.34$\pm$0.27}& 0.128$\pm$0.083 \\
                AV-FacePixels &  8.12$\pm$3.33& 7.59$\pm$2.75& 0.27$\pm$0.24 & 0.116$\pm$0.082\\
                AV-LipPixels& \textbf{8.95$\pm$2.99}& \textbf{8.21$\pm$2.62}& \textbf{0.34$\pm$0.26}&\textbf{0.133$\pm$0.072} \\
                AV-FaceOpt  & 7.24$\pm$4.05 &7.03$\pm$2.90  & 0.23$\pm$0.23& 0.096$\pm$0.099\\
                AV-LipOpt& 8.66$\pm$3.10  & 8.01$\pm$2.64& 0.31$\pm$0.25 &0.127$\pm$0.073\\
                
                AV-FaceMotion& 7.45$\pm$3.61& 7.14$\pm$2.81& 0.23$\pm$0.23 & 0.099$\pm$0.089 \\
                 AV-LipMotion& 8.64$\pm$3.08& 7.98$\pm$2.64& 0.32$\pm$0.26 & 0.127$\pm$0.075 \\
                \bottomrule 
                
                \end{tabular}
                \vspace{-1mm}
                  \caption{}
                 \label{fig:Models3}
          \end{subtable}
          \vspace{-3mm}
            \caption{Mean improvement with standard deviation for unseen speakers with (a) TCD-TIMIT and (b) LRS3 dataset.}
      \end{table*}

\vspace{-4mm}
\subsection{Objective Measures}
\label{ssec:EP}
To evaluate the enhancement performance of different AVSE systems, four different objective metrics are used in our work. The signal-to-distortion ratio (SDR) described in \textit{BSS\_{eval}}\footnote{\url{https://github.com/sigsep/bsseval}} toolkit, measures the distortion present in the enhanced signal, considering  various factors such as remaining interference, newly added artifacts, and channel errors. The scale invariant SDR (SI-SDR) \cite{le2019sdr} is a slightly modified version of SDR, resulting in a simpler, more robust measure. The  perceptual evaluation of speech quality (PESQ) measures the overall perceptual quality and short term objective intelligibility (STOI) represents the correlation with the intelligibility of the signal \cite{michelsanti2020overview}.

\subsection{Results and Discussion}
\label{sec:Result}

The results of our evaluation with unseen speakers in the TCD-TIMIT dataset are presented in Table \ref{fig:Models2}. Due to space constraints and very similar performance of the different features, we only report the SDR and SI-SDR gain in this case. It can be seen that in the considered scenario, the AO-model provides almost no improvement. The AV-Lipembs model performs the best, followed by the AV-FaceOpt and AV-Faceembs model. The other AV models also perform comparatively well. The AV-FaceMotion model achieves the lowest mean SI-SDR gain, and highest standard deviation among all the AV models.

However, it should be noted that the TCD-TIMIT dataset has only a few speakers and very limited vocabulary. The models trained with this dataset also do not achieve consistent performance as can be observed from the standard deviation values.  

In our experiments with the LRS3 dataset (see Table \ref{fig:Models3}), which corresponds to more realistic scenarios, the AV-Faceembs, AV-LipPixels, and AV-Lipembs-LRW models perform similarly, and achieve the best performance. However, we notice a significant performance drop for AV-Lipembs model (the AV-Lipembs performs the best with TCD-TIMIT dataset). This could be associated with the pre-processing system. As mentioned earlier, the lip embedding for this model is achieved using the GRID Corpus dataset, which has a very limited vocabulary. This shows that for better performance of the AVSE system, the underlying Lipreading/VSR models also need to be trained with a varied and larger vocabulary set.

Overall, it is evident from our experiments that the embedding-based visual features perform better than other visual features. However, the time and algorithmic complexity involved in the pre-processing steps to extract the embedding-based features are multiple times higher than the non-embedding-based features. For example, the raw pixels-based feature extraction has a time complexity below 25 ms per frame using the method described in \cite{king2009dlib}, whereas the Facenet-based face embedding extraction process needs over 150 ms using Intel(R) Xeon(R) CPU clocked at 3.60~GHz. It is also worth noting that both the face and lip embeddings require a pre-trianed network, and the performance of the AVSE model depends on the performance of these pre-trained networks. 


Our results show that the optical flow, raw pixels or motion-based features can be suitable alternatives over the embedding-based features for low resource and latency systems. The face-based optical flow or raw pixels features perform better than lip-based features in limited speakers case (see Table \ref{fig:Models2}), but in more practical settings (see Table \ref{fig:Models3}), the lip-based  raw pixel features achieve better performance. Even the AV-LipOpt/AV-LipPixels models perform similarly or better to the face or lip embedding-based models with LRS3 dataset. In case of landmark-based motion features, the lip-based features always outperform the face-based features demonstrating that for AVSE the motion of lip region is much more significant than the whole face region.

The pre-processing system for the optical flow or motion-based features requires a look ahead of one frame, which can increase the latency of the overall system. Also, it can be observed from the performance analysis that in LRS3 dataset, the optical flow-based features result in a slightly degraded performance. This shows that when the consecutive frames contain rapid movements, with the considered 25 FPS sampling frequency for the visual stream, the optical flow-based features may not be able to capture both the temporal and spatial information reliably. For such systems, the raw lip pixels-based features can be used for its low computational complexity and robustness.

\section{Conclusion}
\label{sec:Conclusion}
We empirically investigated the most common visual features used in DNN-based AVSE. Our investigation shows that for low resource systems the use of optical flow or raw pixels-based visual features can be beneficiary and can still achieve similar performance to the embedding-based AVSE models which tend to have a better overall performance at the cost of an involved feature extraction procedure. Through modifications to the visual stream of the base model, we also developed new models for AVSE with non-embedding visual features.

\vfill\pagebreak

\bibliographystyle{IEEEbib}
\bibliography{strings,refs}

\begin{thebibliography}{10}

\bibitem{6853713}
S.~{Braun}, O.~{Thiergart}, and E.~A.~P. {Habets},
\newblock ``Automatic spatial gain control for an informed spatial filter,''
\newblock in {\em ICASSP}, 2014, pp. 830--834.

\bibitem{golumbic2013visual}
E.~Z. Golumbic, Gr.~B. Cogan, C.~E. Schroeder, and D.~Poeppel,
\newblock ``Visual input enhances selective speech envelope tracking in
  auditory cortex at a cocktail party,''
\newblock {\em Journal of Neuroscience}, vol. 33, no. 4, pp. 1417--1426, 2013.

\bibitem{noda2015audio}
K.~Noda, Y.~Yamaguchi, N.~Kazuhiro, H.~G. Okuno, and T.~Ogata,
\newblock ``Audio-visual speech recognition using deep learning,''
\newblock {\em Applied Intelligence}, vol. 42, no. 4, pp. 722--737, 2015.

\bibitem{Chung16}
J.~S. Chung and A.~Zisserman,
\newblock ``Lip reading in the wild,''
\newblock in {\em Asian Conference on Computer Vision}, 2016.

\bibitem{assael2016lipnet}
Y.~M. Assael, B.~Shillingford, S.~Whiteson, and N.~D. Freitas,
\newblock ``Lipnet: Sentence-level lipreading,''
\newblock {\em CoRR}, vol. abs/1611.01599, 2016.

\bibitem{prajwal2020learning}
K.~R. Prajwal, R.~Mukhopadhyay, V.~P. Namboodiri, and C.~V. Jawahar,
\newblock ``Learning individual speaking styles for accurate lip to speech
  synthesis,''
\newblock in {\em CVPR}, 2020, pp. 13796--13805.

\bibitem{chuang2020lite}
S.~Chuang, Y.~Tsao, C.~C. Lo, and H.~M. Wang,
\newblock ``Lite audio-visual speech enhancement,''
\newblock {\em arXiv:2005.11769}, 2020.

\bibitem{hou2018audio}
J.~Hou, S.~Wang, Y.~Lai, Y.~Tsao, H.~Chang, and H.~Wang,
\newblock ``Audio-visual speech enhancement using multimodal deep convolutional
  neural networks,''
\newblock {\em IEEE Transactions on Emerging Topics in Computational
  Intelligence}, vol. 2, no. 2, pp. 117--128, 2018.

\bibitem{ephrat2018looking}
A.~Ephrat, I.~Mosseri, O.~Lang, T.~Dekel, K.Wilson, A.~Hassidim, W.~T. Freeman,
  and M.~Rubinstein,
\newblock ``Looking to listen at the cocktail party: {A} speaker-independent
  audio-visual model for speech separation,''
\newblock {\em SIGGRAPH}, 2018.

\bibitem{afouras2018conversation}
T.~Afouras, J.~S. Chung, and A.~Zisserman,
\newblock ``The conversation: Deep audio-visual speech enhancement,''
\newblock in {\em INTERSPEECH}, 2018.

\bibitem{michelsanti2020overview}
D.~Michelsanti, Z.~Tan, S.g Zhang, Y.~Xu, M.~Yu, D.~Yu, and J.~Jensen,
\newblock ``An overview of deep-learning-based audio-visual speech enhancement
  and separation,''
\newblock {\em arXiv:2008.09586}, 2020.

\bibitem{king2009dlib}
D.~E. King,
\newblock ``Dlib-ml: A machine learning toolkit,''
\newblock {\em The Journal of Machine Learning Research}, vol. 10, pp.
  1755--1758, 2009.

\bibitem{zhang2016joint}
K.~Zhang, Z.~Zhang, Z.~Li, and Y.~Qiao,
\newblock ``Joint face detection and alignment using multitask cascaded
  convolutional networks,''
\newblock {\em IEEE Signal Processing Letters}, vol. 23, no. 10, pp.
  1499--1503, 2016.

\bibitem{schroff2015facenet}
F.~Schroff, D.~Kalenichenko, and J.~Philbin,
\newblock ``Facenet: A unified embedding for face recognition and clustering,''
\newblock in {\em CVPR}, 2015, pp. 815--823.

\bibitem{martinez2020lipreading}
B.~Martinez, P.~Ma, S.~Petridis, and M.~Pantic,
\newblock ``Lipreading using temporal convolutional networks,''
\newblock in {\em ICASSP}, 2020, pp. 6319--6323.

\bibitem{farneback2003two}
G.~Farneb{\"a}ck,
\newblock ``Two-frame motion estimation based on polynomial expansion,''
\newblock in {\em SCIA}. Springer, 2003, pp. 363--370.

\bibitem{simonyan2014two}
K.~Simonyan and A.~Zisserman,
\newblock ``Two-stream convolutional networks for action recognition in
  videos,''
\newblock in {\em Advances in neural information processing systems}, 2014, pp.
  568--576.

\bibitem{baker2004opencv}
S.~Baker and I.~Matthews,
\newblock ``Open\uppercase{CV} dense optical flow,''
\newblock {\em Springer}, vol. 6, pp. 221--255, 2004.

\bibitem{morrone2019face}
G.~Morrone, S.~Bergamaschi, L.~Pasa, L.~Fadiga, V.~Tikhanoff, and L.~Badino,
\newblock ``Face landmark-based speaker-independent audio-visual speech
  enhancement in multi-talker environments,''
\newblock in {\em ICASSP}, 2019, pp. 6900--6904.

\bibitem{gabbay2018seeing}
A.~Gabbay, A.~Ephrat, T.~Halperin, and S.~Peleg,
\newblock ``Seeing through noise: Visually driven speaker separation and
  enhancement,''
\newblock in {\em ICASSP}, 2018, pp. 3051--3055.

\bibitem{harte2015tcd}
N.~Harte and E.~Gillen,
\newblock ``\uppercase{TCD-TIMIT}: An audio-visual corpus of continuous
  speech,''
\newblock {\em IEEE Transactions on Multimedia}, vol. 17, no. 5, pp. 603--615,
  2015.

\bibitem{afouras2018lrs3}
T.~Afouras, J.~S. Chung, and A.~Zisserman,
\newblock ``\uppercase{LRS3-TED}: a large-scale dataset for visual speech
  recognition,''
\newblock {\em arXiv:1809.00496}, 2018.

\bibitem{cooke2006audio}
M.~Cooke, S.~Barker, J.and~Cunningham, and X.~Shao,
\newblock ``An audio-visual corpus for speech perception and automatic speech
  recognition,''
\newblock {\em The Journal of the Acoustical Society of America}, vol. 120, no.
  5, pp. 2421--2424, 2006.

\bibitem{reddy2020interspeech}
C.~K. Reddy, V.~Gopal, E.~Cutler, R.and~Beyrami, R.~Cheng, H.~Dubey,
  S.~Matusevych, R.~Aichner, A.~Aazami, S.~Braun, et~al.,
\newblock ``The interspeech 2020 deep noise suppression challenge: Datasets,
  subjective testing framework, and challenge results,''
\newblock {\em INTERSPEECH}, 2020.

\bibitem{le2019sdr}
J.~L.~Roux, S.~Wisdom, H.~Erdogan, and J.~R. Hershey,
\newblock ``\uppercase{SDR}--half-baked or well done?,''
\newblock in {\em ICASSP}, 2019, pp. 626--630.

\end{thebibliography}

\end{document}